\documentclass[%
 reprint,
 amsmath,amssymb,
 aps,
]{revtex4-2}

\usepackage{xcolor}
\usepackage{graphicx}
\usepackage{dcolumn}
\usepackage{bm}
\usepackage{hyperref}
\usepackage{amsmath}
\usepackage{xspace}
\usepackage{float}
\usepackage{tabularx}
\usepackage{multirow}
\usepackage{ifthen}
\usepackage[caption=false]{subfig}
\usepackage[LGRgreek]{mathastext}

\newcolumntype{C}[1]{>{\centering\let\newline\\\arraybackslash\hspace{0pt}}m{#1}}

\newcommand{\pTone}{$p_T^1$\xspace}

\newcommand{\meanpT}{$\langle p_T \rangle$\xspace}

\newcommand{\akT}{anti-$k_T$\xspace}

\newcommand{\Aj}{$A_{jet}$\xspace}

\newcommand{\GeV}{GeV/$c$\xspace}

\newcommand{\sref}[1]{sec.~\ref{#1}}
\newcommand{\fref}[1]{fig.~\ref{#1}}
\newcommand{\tref}[1]{tab.~\ref{#1}}

\newcommand{\Fref}[1]{Figure~\ref{#1}}
\newcommand{\Tref}[1]{Table~\ref{#1}}

\newcommand{\pp}{$p$+$p$\xspace}

\newcommand{\Pb}{Pb+Pb\xspace}

\newcommand{\sNN}{$\sqrt{s_{\mathrm{NN}}}$\xspace}
\newcommand{\sqrts}{$\sqrt{s}$\xspace}

\newcommand{\pikp}{$\pi^{\pm}$, K$^{\pm}$, p and $\bar{p}$\xspace}

\definecolor{UTOrange}{rgb}{1, 0.51, 0.0}
\newcommand{\BG}{{TennGen}\xspace}
\newcommand{\PYTHIA}{\textsc{PYTHIA}\xspace}
\newcommand{\ptH}{$p_T^{hard min.}$}

\newcommand{\zSub}{$z_{subjet}$\xspace}
\newcommand{\jetW}{$\lambda_1^1$\xspace}

\newcommand{\aCut}{$A>0.6 \pi R^2$\xspace}
\newcommand{\pTCut}{$p_T^1>3.0 GeV/c$\xspace}
\newcommand{\MLpTCut}{$p_T^1>5.036 GeV/c$\xspace}

\extrafloats{200}
\begin{document}


\title{Separating signal from combinatorial jets in a high background environment}

\author{Patrick Steffanic} 
\author{Charles Hughes} 
\author{Christine Nattrass} 
\affiliation{University of Tennessee, Knoxville, TN, USA-37996.}

\begin{abstract}
We study procedures for discriminating combinatorial jets in a high background environment, such as a heavy ion collision, from signal jets arising from a hard-scattering. We investigate a population of jets clustered from a combined \PYTHIA+\BG event, focusing on jets which can unambiguously be classified as signal or combinatorial jets. By selecting jets based on their kinematic properties, we investigate whether it is possible to separate signal and combinatorial jets without biasing the signal population significantly.  We find that, after a loose selection on the jet area, surviving combinatorial jets are dominantly imposters, combinatorial jets with properties indistinguishable from signal jets. We also find that, after a loose selection on the leading hadron momentum, surviving combinatorial jets are still dominantly imposters. We use rule extraction, a machine learning technique, to extract an optimal kinematic selection from a random forest trained on our population of jets. In general, this technique found a stricter kinematic selection on the jet's leading hadron momentum to be optimal. We find that it is possible to suppress combinatorial jets significantly using this machine learning based selection, but that some signal is removed as well.  Due to this stricter kinematic selection, we find that the surviving signal is biased towards quark-like jets. Since similar selections are used in many measurements, this indicates that those measurements are biased towards quark-like jets as well.  These studies should motivate an increased emphasis on assumptions made when suppressing and subtracting combinatorial background and the biases introduced by methods for doing so.
\end{abstract}

\maketitle

\section{Introduction}

A hot, dense, strongly interacting liquid of quarks and gluons called the Quark Gluon Plasma (QGP) is briefly created in high energy heavy ion collisions~\cite{Adcox:2004mh,Adams:2005dq,Back:2004je,Arsene:2004fa}.  Two of the key signatures of the formation of the QGP are jet quenching and hydrodynamical flow. There have been extensive measurements of jets in a QGP, including single particle spectra, jet spectra, fragmentation functions, and jet substructure~\cite{Connors:2017ptx}.  While there have been some constraints on the properties of the medium due from measurements of jets~\cite{Burke:2013yra, JETSCAPE:2021ehl}, the era of quantitative measurements of jets is just beginning.  A detailed understanding of jet quenching requires a quantitative understanding of the background.

The paradigm the field uses to separate signal and background assumes that particles are either from jets or from background processes. While this approach is commonly used in heavy ion collisions, we note that similar approaches to a combinatorial background may be applicable to \pp collisions in a high pile up environment.  The signal is assumed to be a cluster of particles from a hard process.  The background includes particles from all other processes, including other jets.  Many background particles are correlated, either due to resonances or flow.  A jet finder will cluster all particles into a jet, so the background impacts the signal both through background particles clustered with signal jets and jets consisting exclusively of background particles, called combinatorial jets.

In practice, there is some ambiguity in whether particles are from the signal or the background.  When partons interact with the medium, they may leave some of their energy and momentum in medium particles, for instance through a wake from movement of the parton through a fluid~\cite{Yang:2022nei}.  It is unclear in that case if those particles should be considered signal or background.  Furthermore, at low momenta, the definition of a jet itself becomes ambiguous, as there is no clear threshold for when a process is hard enough to result in a jet and a jet finder will cluster particles with correlated momenta independent of their production mechanism.

The assumption that particles are either from the signal or from the background works well for describing combinatorial jets, jets composed entirely of particles from the background. This can be seen by the agreement between the sum of the energies of all particles found in a random cone in a heavy ion collision and expectations for drawing random particles with momenta matching that observed in the data~\cite{Abelev2012}.  While there are some deviations between these observations and expectations from a random sample, there is even good agreement with PYTHIA Angantyr, which includes correlations from resonances and mini-jets~\cite{Hughes:2020lmo}.  The contribution from combinatorial jets is typically managed by focusing on high momentum jets or suppressing the background through kinematic selections such as requiring a minimum momentum for the highest momentum particle~\cite{ALICE:2019qyj}.  Combinatorial jets have also been described by mixed events and their contribution subtracted when their contribution is limited by a coincidence with a high momentum hadron $180^{\circ}$ away~\cite{Adamczyk:2017yhe}.

The contribution of background particles in jets has typically been subtracted either through an iterative procedure to estimate background contributions ~\cite{ATLAS:2012tjt,CMS:2016uxf}, or by estimating the background per unit area~\cite{ALICE:2019qyj,Adamczyk:2017yhe}.  Recently a shallow neural network was applied to estimate the background contribution to jet measurements to extend these measurements to lower momenta and larger R~\cite{Haake:2018hqn,Bossi:2020hwt}.

Measurements of jets in heavy ion collisions are limited at low momenta both because the number combinatorial jets becomes comparable to the number of signal jets and because measurements have large uncertainties when fluctuations in the background contributions to signal jets are comparable to the jet momenta.  In this study, we use a model with a randomly generated background from TennGen~\cite{Hughes:2020lmo}, described in \sref{Sec:TennGen}, and a signal from PYTHIA, described in \sref{Sec:PYTHIA}.  We define unambiguous samples of signal and combinatorial jets in \sref{sec:combJets}, and characterize them by their properties, summarized in \sref{Sec:Observables}.  We introduce the silhouette value in \sref{Sec:Silhouette} as an alternate means to look at whether or not signal and combinatorial jets are distinguishable.  In \sref{Sec:MachineLearning} we describe a machine learning approach to optimizing the separation of signal and combinatorial jets and in \sref{Sec:Zsub} we describe the leading subjet fraction, which we use to look for bias introduced by the kinematic selections used to separate signal and combinatorial jets.  In \sref{Sec:results} we describe the kinematic selections identified, discuss the limitations in separating signal and combinatorial jets, and investigate possible biases imposed by these kinematic selections.

\section{Method}
We investigate ways to distinguish signal and combinatorial jets in heavy-ion collisions. We use a model to unambiguously define signal and combinatorial jets, which is not possible in experiment.  We use machine learning to explore promising observables, and their associated kinematic selections, that may have been overlooked. We consider jet observables, described in \sref{Sec:Observables}, whose interpretation is comparable in experiment and theory. We demonstrate that the machine learning system reproduces the standard kinematic selections. Furthermore, we show that any kinematic selection admits combinatorial jets whose properties are indistinguishable from signal jets. To further reduce the number of combinatorial jets, the number of signal jets is necessarily reduced, introducing a bias.

We use \BG~\cite{Hughes:2020lmo,TennGenGITHUB}, briefly described in \sref{Sec:TennGen}, for a realistic background \Pb event at \sNN = 2.76 TeV with correlations due to flow but no other physics correlations.  
We embed a \pp  collision produced with \PYTHIA 6 \cite{Sjostrand:2006za} using the Perugia 2011c\cite{Skands:2010ak} tune at \sqrts = 2.76 TeV, briefly described in \sref{Sec:PYTHIA}, in the \BG event to generate a jet signal. We cluster the combined event with the \akT jet finder, producing a population of jets with particles from both \PYTHIA and \BG.  We can then classify jets as combinatorial or signal by how much momentum is from \BG and \PYTHIA particles, as described in \sref{sec:combJets}.

We use a number of jet observables, discussed in \sref{Sec:Observables}, to characterize our jet population and evaluate the effect of different kinematic selections on signal and combinatorial jets. 
We introduce silhouette values, used to quantify how similar an object is to other object in its cluster, in \sref{Sec:Silhouette} and use the distribution of silhouette values to evaluate the similarity between signal and combinatorial jets.  We use a random forest, a type of machine learning algorithm, coupled with a decision tree to search for optimal selections that distinguish between signal and combinatorial jets.  This algorithm is described in \sref{Sec:MachineLearning}.
We study the momentum fraction carried by the leading subjet, described in \sref{Sec:Zsub}, to understand the bias towards quark-like jets imposed by kinematic selections.

\subsection{Background generation}\label{Sec:TennGen}
\BG emulates a realistic background for jet studies in heavy ion collisions by throwing random particles which match the multiplicities~\cite{Abelev:2013vea}, momentum distributions~\cite{Abelev:2012hxa}, and azimuthal distributions~\cite{Aad:2014fla} of single particles.  \BG is described in greater detail in~\cite{Hughes:2020lmo} and the source code is publicly available~\cite{TennGenGITHUB}.

The measured single particle double differential spectra for \pikp from~\cite{Abelev:2012hxa} are fit to a Boltzmann-Gibbs Blast Wave distribution~\cite{Ristea:2013ara,Schnedermann:1993ws}.  This distribution is used to randomly select a momentum for each particle.  The single particle flow coefficients~\cite{Aad:2014fla} are used to determine the azimuthal distribution of particles with that momentum, and a random azimuthal angle is determined from that distribution.  The pseudorapidity $\eta$ of the particle is randomly selected from a flat distribution in $-0.9 < \eta < 0.9$ to match the $\eta$ acceptance of the ALICE detector.
The multiplicity of each particle species is determined from ratios~\cite{Abelev:2013vea} and is scaled up assuming a constant charged particle multiplicity per unit pseudorapidity. We use 0–5\% central \BG events, which have $\frac{dN}{d\eta}$ = 2217 and $\langle p_T \rangle$ = 0.56 GeV/c. The multiplicities are determined from measurements of the charged particle multiplicities in~\cite{Aamodt:1313050}.

By construction, \BG events contain no correlations other than those due to flow. Previous studies indicate that the background can be described by randomly selecting particles from the single particle distribution~\cite{Abelev2012,Adamczyk:2017yhe,Hughes:2020lmo}. Some contributions from mini-jets and resonances could arguably be considered signal and we aim to study the properties of jets which are unambiguously combinatorial.  The absence of these additional correlations in TennGen is therefore a desirable feature for these studies. As such, all particles in \BG are considered background particles for measurements of jets.

\subsection{Signal generation}\label{Sec:PYTHIA}
\PYTHIA ~\cite{Sjostrand:2006za} is a general purpose Monte Carlo event generator. In this study we generate proton-proton events at \sqrts = 2.76 TeV in \PYTHIA using the Perugia 2011c\cite{Skands:2010ak} tune and require a hard scattering with various  \ptH. Other parameters are left at their defaults.
Jets are clustered from primary $\pi^0$, \pikp particles with $|\eta|$ $<$ 0.9 using the \akT algorithm implemented in FastJet v3.2.1~\cite{Cacciari_2012} with various jet resolution parameters $R$. We keep \PYTHIA events which have one jet with at least 80\% of the \ptH within the acceptance in pseudorapidity and embed them in a \BG event.  This leads to a slightly different distribution of jets than if we were to use minimum bias \PYTHIA events but does not qualitatively change the conclusions. Particles from \PYTHIA are considered to be signal particles.  There is a small underlying event in \pp, but we treat this as negligible compared to the jet signal.

\subsection{Signal and combinatorial jets}\label{sec:combJets}

We cluster all the particles from the combined \BG and \PYTHIA event with the \akT jet finder. We calculate the fraction of each jet's momentum carried by \PYTHIA particles.  Jets which have less than $2 \pi R^2$ \GeV from \PYTHIA are classified as combinatorial jets.  This allows up to the average momentum in a random cone from the underlying event in \PYTHIA assuming an average momentum density of 2 \GeV per unit area.  Jets which contain at least 0.8 \ptH \GeV from \PYTHIA particles are classified as signal jets. This ensures that only PYTHIA jets are classified as signal jets, without any combinatorial jets classified incorrectly as signal jets.  The remainder are not classified.  This allows us to identify unambiguous samples of signal and combinatorial jets. 

The results presented in \sref{Sec:results} focus on $R=0.5$ and \ptH = 40 \GeV and results for additional resolution parameters ranging from R=0.2 to 0.6 for \ptH ranging from 10 to 80 \GeV are given in the Supplemental Material ~\cite{[{See Supplemental Material at }][{ for kinematic selections of other values of R, and \ptH.}]supp}.

\subsection{Observables}\label{Sec:Observables}

Our objective is to identify observables which may help discriminate between signal and combinatorial jets, could realistically be used in data, and either would lead to a negligible bias in the surviving population of signal jets or whose bias could be reproduced well in model calculations. In particular, our aim is to study low momentum jets to investigate approaches to decreasing the lower threshold for jet measurements and decrease the systematic uncertainties associated with background subtraction in this region. We started with observables which could be measured on an individual jet basis and eliminated observables using a combination of our knowledge of the strengths and weaknesses of these observables and statistical techniques. We only used observables which are reliably measurable in data and calculable in models with reasonable uncertainties. This excludes the $n^{th}$ leading particle's momentum for n$>$1, for instance, because they would be difficult to model accurately, and may make the model sensitive to fluctuations in data and models. We used the scikit learn implementation of principle component analysis~\cite{scikitPCA} to better understand which observables are redundant, and feature importance from our random forest~\cite{RandomForests} to remove observables which had little discriminatory power. 
The observables we chose are summarized in \tref{tab:features}. The principle component analysis indicates that these observables explain 98\% of the variance in our data. Of that, \pTone accounts for the majority of the variance. While a similar approach may lead to a different set of observables, we think that this is a realistic set of observables which could be used to discriminate between signal and combinatorial jets.

\renewcommand{\arraystretch}{2}
\begin{table*}[t]
    \centering
    \begin{tabular}{|ccc|}
\hline
Symbol & Name & Definition \\
\hline
\Aj & Jet Area & Area covered by all jets constituents \\
\pTone & Leading Hadron Momentum & Momentum of leading jet constituent \\ 
\jetW & Jet Width & $\sum_{i=1}^{N_{constit.}} z_{consit., i}\cdot\Delta R_{constit. i, jet}$\\
\meanpT & Mean $p_T$ & $\frac{1}{N_{constit.}}\sum_{i=1}^{N_{constit.}} p_{T,i}$\\ \hline
\end{tabular}
    \caption{Observables used to characterize each jet population.}
    \label{tab:features}
\end{table*}
\renewcommand{\arraystretch}{1}

\subsubsection{Area}
To calculate the area of jet, we add many very soft particles("ghosts") to the event, counting how many ghosts are clustered into our jet. The jet area is given by
\begin{equation}
    A_{jet} = A_g \langle N_g \rangle
\end{equation}

\noindent where $A_g$ is the area of a single ghost and $\langle N_g \rangle$ is the average number of ghosts clustered into our jet~\cite{Cacciari_2008}.

\subsubsection{Leading hadron momentum}
The leading hadron momentum, \pTone, is the highest momentum jet constituent.  While there is some model dependence in calculating a leading hadron momentum in theory calculations because hadronization is non-perturbative, such calculations are fairly robust since single hadron observables in \pp collisions agree well with pQCD calculations~\cite{Han:2022zxn}.  We opted not to include the subleading hadron momentum and beyond because they would make model calculations more sensitive to details of hadronization.

\subsubsection{Jet width}
The jet width\cite{Reichelt:2021svh}, \jetW, is given by
\begin{equation}
    \lambda_1^1 = \sum_{i=1}^{N} z_{i}\cdot \frac{\Delta R_{i, jet}}{R}
\end{equation}
\noindent where $N$ is the number of constituents in the jet, $z_{i}$ is the momentum fraction carried by constituent $i$, and $\Delta R_{i, jet}$ is the distance in $\eta$-$\phi$ space between constituent $i$ and the jet axis.  This provides a measure of how far constituents are from the jet axis on average.

\subsubsection{Mean constituent transverse momentum}
Mean constituent transverse momentum including background, \meanpT, is the average $p_T$ of the jet's constituents. \BG is expected to reproduce the \meanpT from data fairly accurately. 

We investigated the use of higher order moments of the momentum distribution, such as the standard deviation. These features provide useful information, but reproducing them in a model would require getting the single particle spectrum and the fragmentation functions correct to high precision. We concluded this would add too much model dependence.

\subsection{Silhouette values}\label{Sec:Silhouette}

We borrow a technique from data science to characterize the overlap between our different populations. Our two populations are unambiguous signal, and combinatorial jets.  The silhouette value\cite{ROUSSEEUW198753} describes, for each signal or combinatorial jet, whether it shares more characteristics with its own cluster or the other cluster.  The silhouette value for the jet with index i is given by
\begin{equation}
    S(i) = \frac{b(i) - a(i)}{max\{a(i), b(i)\}}
\end{equation}

\noindent$a(i)$ is the mean in-class distance,
\begin{equation}
    a(i) = \frac{1}{N_{I}-1} \sum_{j\neq i}^{N_I} d(i, j)
\end{equation}
\noindent where the sum runs over jets of the same class as jet i, and $N_{I}$ is the number of jets in the same class as jet i; $b(i)$ is the mean out-of-class distance,
 \begin{equation}
     b(i) = \frac{1}{N_{J}} \sum_{j}^{N_J} d(i, j)
 \end{equation}
  \noindent where the sum here runs over jets of the opposite class from jet i, and $N_{J}$ is the number of jets in the opposite class. The $d(i,j)$ in both equations represents the distance between jet i and jet j. This is typically the Euclidean distance computed in feature space. We standardize our data so that each feature lies in the range [0,1], ensuring that each feature contributes equally to the silhouette value, rather than features with larger ranges having a larger effect.
 \begin{multline}
     d(i,j) = \biggl\{\bigg(\frac{A_{jet,j}-A_{jet,i}}{A_{jet}^{max} - A_{jet}^{min}}\bigg)^2+\bigg(\frac{p_{T,j}^1-p_{T,i}^1}{p_{T}^{1,max}-p_{T}^{1,min}}\bigg)^2\\+\bigg(\frac{\lambda_{1,j}^1-\lambda_{1,i}^1}{\lambda_{1}^{1,max}-\lambda_{1}^{1,min}}\bigg)^2+\bigg(\frac{\langle p_T \rangle_j - \langle p_T \rangle_i}{\langle p_T \rangle^{max} - \langle p_T \rangle^{min}}\bigg)^2 \biggr\} ^\frac{1}{2}.
 \end{multline}
 
  Silhouette values have a range from -1 to 1. A jet with a positive silhouette value is more similar to its own class than the other class. A negative silhouette value indicates the opposite, the jet is more similar to jets from the other class than to jets from its own class.  While it would be possible to extend the list of observables, we do not anticipate others would have any significant distinguishing power.
We rejected a number of interesting observables that were either too complex to faithfully simulate or too difficult to measure in data. Silhouette values could help determine the effectiveness of any new observable to distinguish signal and combinatorial jets.

\subsection{Kinematic selection optimization}\label{Sec:MachineLearning}

We train a machine learning system to learn about the relationship between a jet's features and whether it is a signal jet or a combinatorial jet using the input features described in \sref{Sec:Observables}. We use a random forest, but a neural network or any other algorithm capable of classification could be used. We then train a decision tree on the predictions of the random forest.  The decision tree is a classification algorithm that uses a series of kinematic selections to classify each jet. This technique is a type of rule extraction called the oracle method\cite{johansson2006not}. We extract the top-level node of our trained decision tree, the kinematic selection which best separates signal and combinatorial jets. We then evaluate the biases and background reduction of any such selection. Our goal is to extract the useful information from machine learning while avoiding its downsides, namely over-fitting and model dependence. The details of our specific machine learning system are mentioned below.

\subsubsection{Decision trees}

We use the sci-kit learn implementation of the decision tree\cite{scikitDT}, which is based on the Classification and Regression Trees algorithm described in~\cite{Breiman1984}. A decision tree recursively partitions the feature space such that the samples with the same labels are grouped together. Each candidate split, $\theta$, consists of a feature and a threshold; in our case, a kinematic selection.  The selection is evaluated according to an impurity function H and its quality is determined by 
\begin{equation}
    G(Q_m, \theta) = \frac{N_m^{left}}{N_m}H(Q_m^{left}(\theta)) + \frac{N_m^{right}}{N_m}H(Q_m^{right}(\theta))
\end{equation}

\noindent where $Q_m$ are the data at node m, $N_m$ is the total number of samples at node m, and the left (right) superscripts indicate that the data are below (above) the threshold. The algorithm selects the split which minimizes G. There are two primary choices for the impurity function H, Gini impurity
\begin{equation}
     \sum_{i=1}^{N_c} p_i(1-p_i)
\end{equation}
and entropy
\begin{equation}
     -\sum_{i=1}^{N_c} p_i log_2 p_i
\end{equation}

\noindent where $N_c$ is the number of classes and $p_i$ is the probability of randomly picking an element of class i. We use Gini impurity in this study.

Decision trees can be prone to over-fitting the data, resulting in poor generalization. Additionally, their accuracy is quite dependent on appropriate hyper-parameter tuning~\cite{scikitDT}. We used scikit-learn v1.2.0 and the default parameters of the DecisionTreeClassifier except for the max\_depth parameter, which was set to 3.

\subsubsection{Random forests}

In order to overcome the problem of over-fitting the random forest algorithm\cite{breiman2001random} is used. 
This is an ensemble method involving training hundreds of randomized decision trees and averaging their predictions for a more robust, general prediction. There are two sources of randomness in the algorithm: each decision tree sees only a random sample of the data, and at each splitting the decision tree can use either all of the features or a random subset of a chosen size. Each decision tree is trained independently from the others. The variance of a random forest is smaller than that of a single decision tree due to the injection of randomness and resulting average over the decision trees. The prediction for each jet becomes the ground truth for the "oracle" decision tree. It should be noted that the scikit-learn implementation is not exactly as in~\cite{breiman2001random}, as described in~\cite{scikitRF}. We used scikit-learn v1.2.0 and the default parameters of the RandomForestClassifier except as noted in \tref{tab:RFParams}.

\begin{table}[]
    \centering
    \begin{tabular}{|ccc|}
    \hline
    Parameter name & This study & Default \\
    \hline
       n\_estimators  & 200 & 100 \\
       max\_depth & 3 & None \\
       min\_samples\_leaf & 100 & 1 \\
       min\_weight\_fraction\_leaf & 0.1 & 0.0 \\
       max\_samples & 0.9 & 1.0 \\
       random\_state & 42 & None \\
       \hline
    \end{tabular}
    \caption{Non-default parameters used for the RandomForestClassifier in this study.}
    \label{tab:RFParams}
\end{table}

After the "oracle" decision tree is trained, we extract a kinematic selection--the top node of the tree--as well as the signal jet and combinatorial jet rejection rates after applying the selection. We then apply these selections to our data and study the potential biases that they introduce to determine their usefulness in analyses. 

\subsection{Leading sub-jet momentum fraction}\label{Sec:Zsub}
The concept of quark or gluon jets is only defined at leading order, so any realistic distinction, even in a model calculation, is somewhat ad hoc.  However, models predict differences between quark-like and gluon-like jets. Suppression of background by selecting on the kinematic properties of jets may bias the population of surviving jets towards or away from quark-like or gluon-like jets because  quark-like jets fragment into fewer, harder particles which are closer on average to the jet axis than gluon-like jets~\cite{Abreu:1995hp,OPAL:1995ab}.  The observables in \sref{Sec:Observables} are therefore expected to have different distributions for quark-like and gluon-like jets.        

We use the leading subjet momentum fraction to qualitatively evaluate whether kinematic selections impose a substantial bias towards quark-like jets. The jet constituents are reclustered with the \akT jet finder with a smaller radius parameter, $R=0.1$.  The leading subjet momentum fraction, \zSub, is the fraction of the jet's total momentum in the leading subjet~\cite{Apolin_rio_2018}
\begin{equation}
    z_{subjet} = \frac{p_T^{subjet}}{p_T^{jet}}.
\end{equation}
Studies of this observable indicate that quark-like jets have a higher \zSub while gluon-like jets have a lower \zSub~\cite{Neill:2021std,MateuszPrivate}.  We use only PYTHIA particles to determine \zSub in our sample.  We note that the connection between \zSub and the leading parton is more tenuous for smaller $R$ and lower momenta, but nevertheless consider this a reasonable observable to test for biases towards quark-like jets.

\section{Results}\label{Sec:results}

\begin{figure*}
    \centering
    \includegraphics[width=\linewidth]{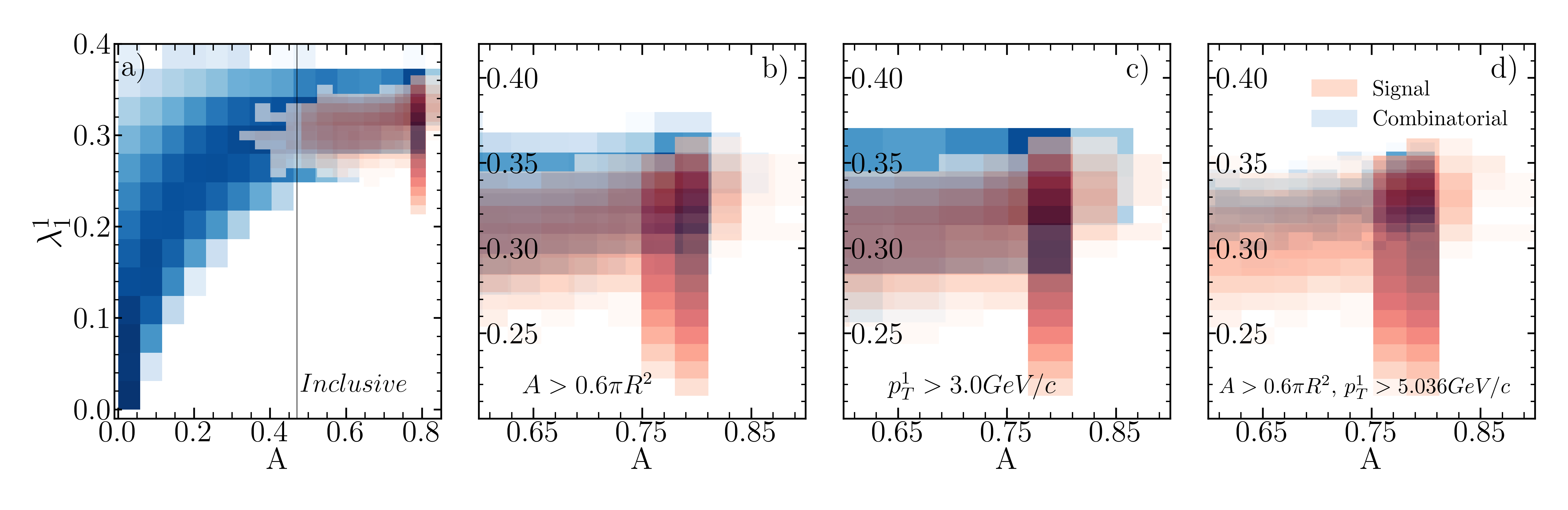}
    \caption{Distribution of \Aj vs. \jetW for signal and combinatorial jets with $R=0.5$ and \ptH$=40$ \GeV after each kinematic selection. The z-axis has a log scale.
    }
    \label{fig:blob_plot}
\end{figure*}

\newcommand{\isArxiv}{True}
\newcommand{\arXivLink}{arXiv link to be added}
\ifthenelse{\equal{\isArxiv}{True}}{
\newcommand{\result}{Results from other selections can be found in the appendices. }
}
{
\newcommand{\result}{Results for other values of R, and \ptH can be found in Supplemental Material~\cite{[{See Supplemental Material at }][{ for kinematic selections of other values of R, and \ptH.}]supp}.
}}

We successively apply four different kinematic selections. Here we present results for $R=0.5$ and \ptH$=40$ \GeV. \result    \Fref{fig:blob_plot}(a) shows the inclusive distribution of jet area versus \jetW for both signal and combinatorial jets.  This shows the overlap between both jet populations. There are few signal jets at small areas, indicating that the large area region can be selected with little bias. We therefore apply the ALICE selection of $A>0.6 \pi R^2$. For jets with resolution parameters $R>0.3$ and \ptH$<30$ \GeV there may be some bias imposed due to more signal jets in the low area region, shown in Supplemental Material. 
\Fref{fig:area_cut_effect} shows the distribution of \pTone, \jetW, and \meanpT for signal and combinatorial jets before and after this selection normalized by the total number of jets before the selection. Changes in the signal distributions are negligible except for resolution parameters R=0.5, and 0.6 when \ptH $<$ 20 \GeV, shown in Supplemental Material, while the surviving combinatorial jet distributions for \jetW and \meanpT become more like those of signal jets.  We call these combinatorial jets which look like signal jets "imposter jets."  The selection on area is so effective because many combinatorial jets consist of a single particle; such jets have small areas with low \jetW and a \meanpT distribution closer to the inclusive particle momentum distribution.

\begin{figure*}
    \centering
    \includegraphics[width=\linewidth]{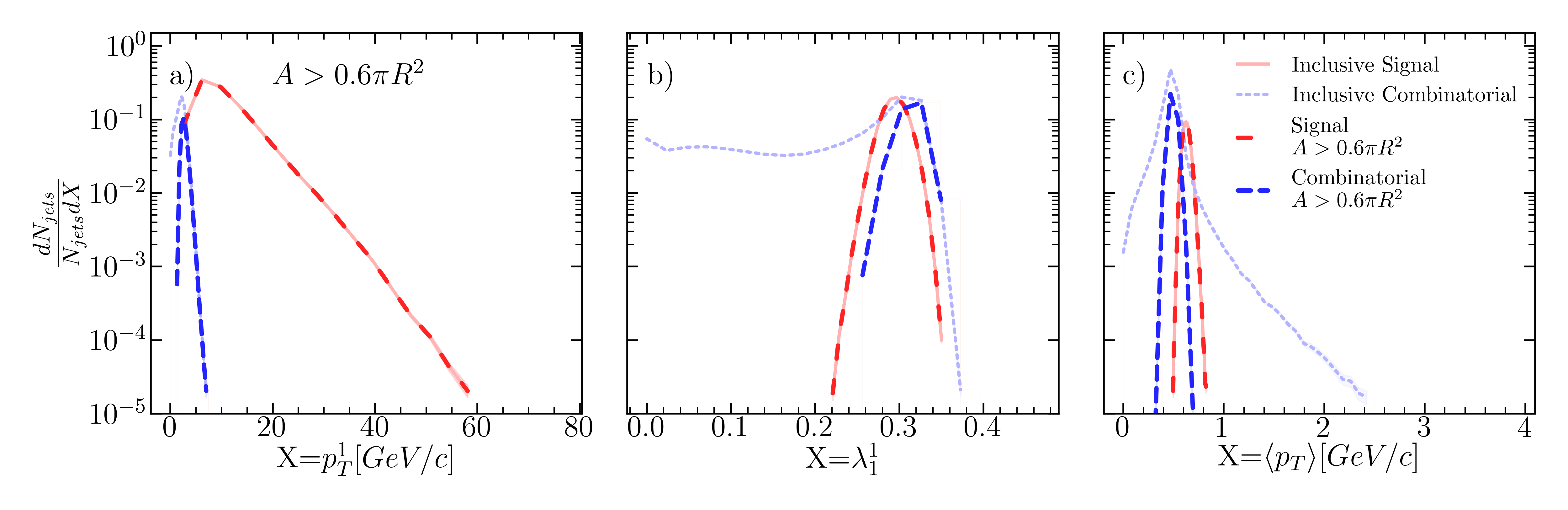}
    \caption{ Distributions of leading hadron momentum, jet width, and mean constituent momentum for jets with $R=0.5$ and \ptH$=40$ \GeV before and after excluding jets with $A<$ $0.6*\pi R^2$. The overall reduction of combinatorial jets was 66.3\%.}
    \label{fig:area_cut_effect}
\end{figure*}

The jets remaining after applying a kinematic selection in \fref{fig:area_cut_effect} indicate a significant difference between \pTone distributions for signal and combinatorial jets, while the \jetW distributions overlap significantly for signal and combinatorial jets. The differences for both \pTone and \jetW increase with \ptH and decrease with increasing resolution parameter.  The \meanpT distributions indicate a separation between signal and combinatorial jets. They follow the same trends in \ptH and R as in \pTone and \jetW, but the differences are not as large as those in the \pTone distributions. Furthermore, the \meanpT distribution would be difficult to recreate in simulation, meaning that any biases would be difficult to reproduce for comparisons between models and data. \Fref{fig:blob_plot}(b) shows that after this selection, the distributions of signal and background jets overlap significantly, so any further selection will suppress signal jets as well.

\begin{figure*}
    \centering
    \includegraphics[width=\linewidth]{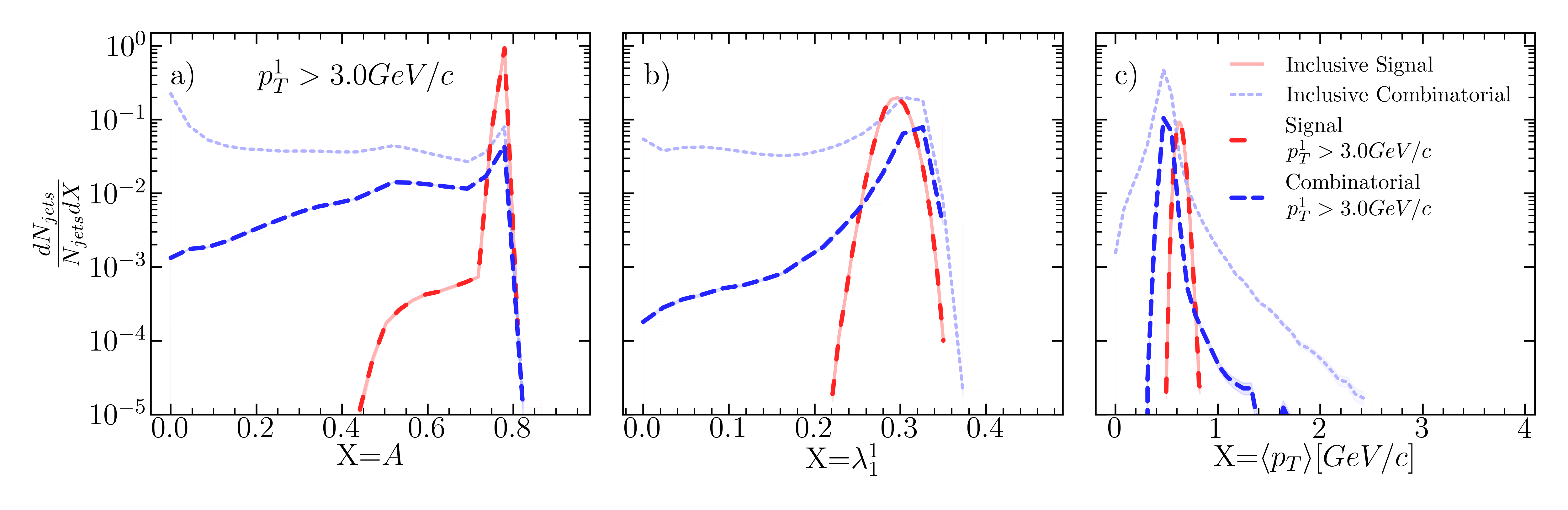}
    \caption{ Distributions of jet area, jet width, and mean constituent momentum for jets with $R=0.5$ and \ptH$=40$ \GeV before and after excluding jets with $p_T^1<$ 3 \GeV. The overall reduction of combinatorial jets was 81.6\%.}
    \label{fig:lead_had_cut_effect}
\end{figure*}

We then investigate a selection of $p_T^1> 3.0 $ \GeV.  The distributions of $A$, \jetW, and \meanpT for surviving signal and combinatorial jets are shown in \fref{fig:lead_had_cut_effect}.  As for the selection on area, there is no apparent difference in the distributions for surviving signal jets.  Surviving combinatorial jets are imposters. \Fref{fig:blob_plot}(c) shows that the properties of the surviving jets, indeed, still overlap significantly. A selection on \pTone is effective at suppressing background, but it is not collinear safe and may introduce biases.  Biases may be unavoidable, and the impact of a selection on \pTone is at least likely to be easier to reproduce in model calculations than other observables, such as \meanpT.

\begin{figure*}

    \includegraphics[width=\linewidth]{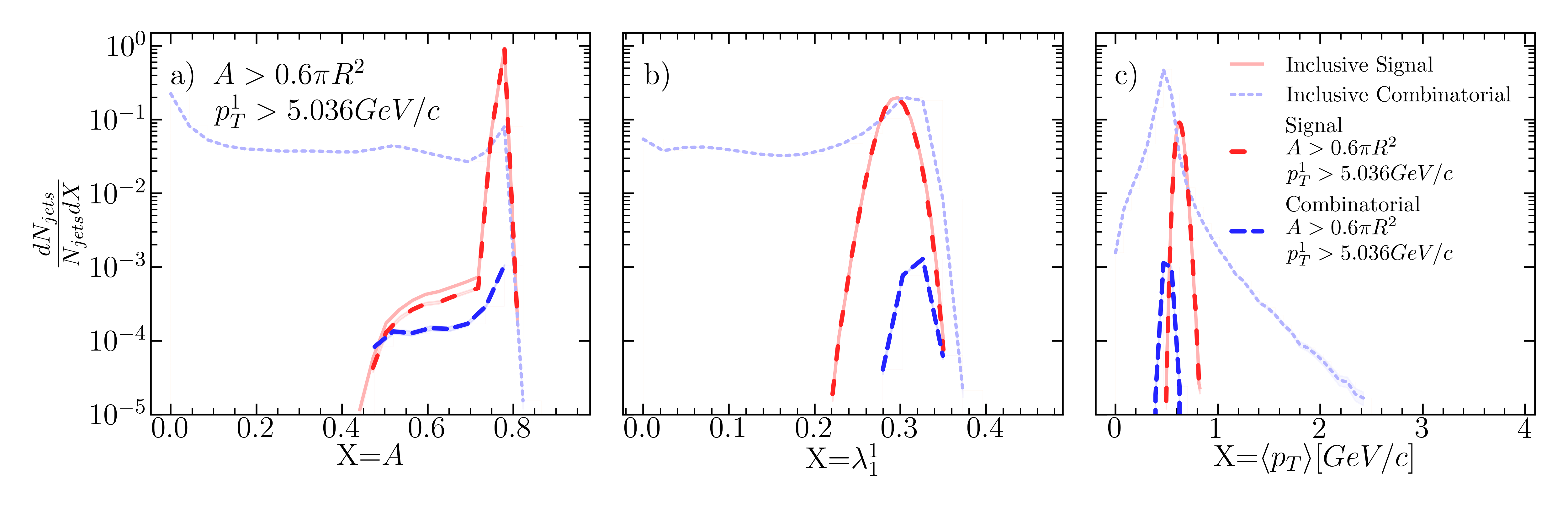}    \caption{ Distributions of jet area, leading hadron momentum, jet width, and mean constituent momentum for jets with $R=0.5$ and \ptH$=40$ \GeV before and after excluding jets with $A<$ $0.6*\pi R^2$, and $p_T^1 < 5.036$ \GeV. The overall reduction of combinatorial jets was 99.8\%. This tighter selection rejects predominantly gluon-like signal jets, as shown in Figure ~\ref{fig:z_subjet}.
    }
    \label{fig:ml_cut_effect}

\end{figure*}

\ifthenelse{\equal{\isArxiv}{True}}{
\newcommand{\refAppTab}{ given in \tref{tab:MLcuts}.}
}{
\newcommand{\refAppTab}{ given in Suppemental Material~\cite{[{See Supplemental Material at }][{ for kinematic selections of other values of R, and \ptH.}]supp}}
}

We then approach kinematic selections using rule extraction from a random forest.  The selection on area is efficient, cutting little or no signal while eliminating significant background, but a tighter selection which eliminated significant signal might be difficult to reproduce in models.  We therefore keep the area selection and use the random forest, as well as the decision tree, both described in \sref{Sec:MachineLearning}, to identify the best kinematic selection among \pTone, \jetW and \meanpT.  The algorithm found that the optimal selection was \pTone$> 5.036 $ \GeV for jets with R=0.5, and \ptH=40 \GeV. For other values of resolution parameter and \ptH the trend is generally a tighter selection on \pTone ranging from 3.5 \GeV to 5.2 \GeV,\refAppTab. The algorithm finds \meanpT to be the optimal selection for \ptH $> 60$ \GeV, but we do not explore it in this study for the reasons mentioned above. The extracted selections reproduce the machine learning systems predictions within 1\% accuracy. We can see in \fref{fig:blob_plot}(d) that the remaining combinatorial jets are imposters.  \Fref{fig:ml_cut_effect} shows the impact of this selection on the distribution of $A$, \jetW, and \meanpT for signal and combinatorial jets.  The suppression of combinatorial jets is much greater than that seen in \fref{fig:area_cut_effect} or \fref{fig:lead_had_cut_effect}.

\begin{figure*}
    \centering
    \includegraphics[width=\linewidth]{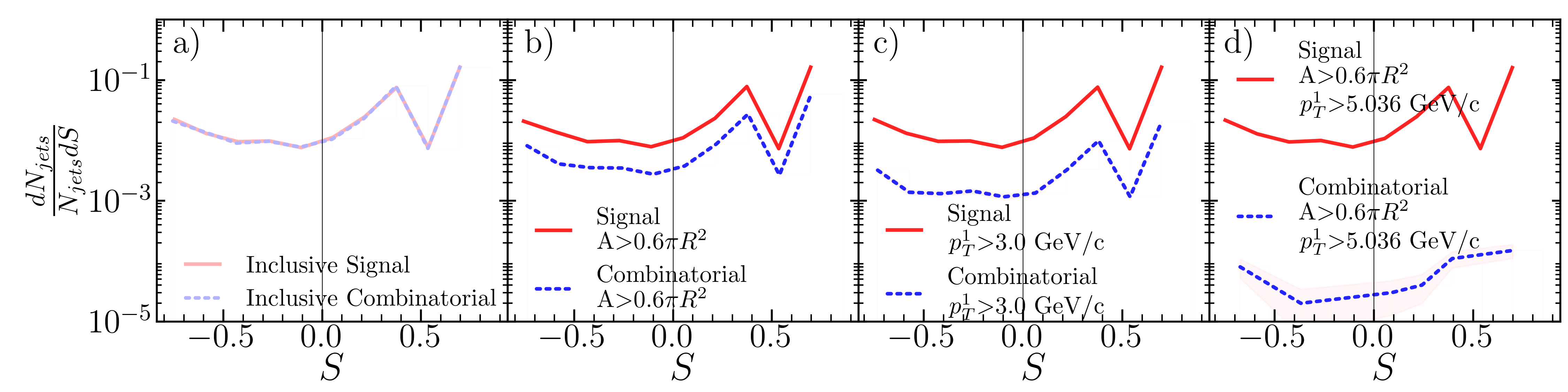}
    \caption{Distributions of silhouette values for signal and combinatorial jets with R=0.5, \ptH=40 \GeV after each kinematic selection. The silhouette values are recomputed for each new population of jets. Each population is normalized by the total inclusive number of jets. The silhouette value describes whether a jet is more similar to its own class, or to the opposite class as described in \sref{Sec:Silhouette}.}
    \label{fig:silhouette_values_selection}
\end{figure*}

\begin{table}
\begin{tabular}{|cccc|}
\hline
   Selection & Population & $\%>0$ & $\%<0$ \\
   \hline
    \multirow{2}{*}{Inclusive} & Signal & 100\% & 0\% \\
     & Combinatorial & 63\% & 37\% \\
    
    \multirow{2}{*}{\aCut} & Signal & 84\% & 16\% \\
     & Combinatorial & 84\% & 16\% \\
    
    \multirow{2}{*}{\pTCut} & Signal & 84\% & 16\% \\
     & Combinatorial & 84\% & 16\% \\
    
    \aCut,   & Signal & 84\% & 16\% \\
     \MLpTCut & Combinatorial & 84\% & 16\% \\
    \hline
\end{tabular}
\caption{The percentage of R=0.5, \ptH=40 \GeV jets which have silhouette values above and below zero.}
\label{tab:ss_val_over_under}
\end{table}

We have only considered rectilinear selections so far, but it is possible that some combination of observables described in \sref{Sec:Observables} may be more effective at distinguishing signal and combinatorial jets.  The silhouette values described in \sref{Sec:Silhouette} are designed to help determine whether two sets are distinguishable or whether there is too much overlap.  \Fref{fig:silhouette_values_selection} shows the silhouette values for each of our selections.  \Tref{tab:ss_val_over_under} lists the fraction of signal and combinatorial jets above and below zero.  The inclusive distribution in \fref{fig:silhouette_values_selection}(a) shows that, before the area selection, there is a large population of combinatorial jets with a positive silhouette value (63\%).  These combinatorial jets are easier to distinguish from signal jets.  However, there is also a significant population of imposters, those with a negative silhouette value (37\%).  \Fref{fig:silhouette_values_selection}(b), (c), and (d) show the silhouette values for signal and combinatorial jets after the area selection, \pTone selection, and the selection chosen by the random forest.  In these cases, the distributions of silhouette values for signal and combinatorial jets have similar shapes, indicating that either is just as likely to look like its own group or like the other group.  

\begin{figure}
    \centering

    \includegraphics[width=\columnwidth]{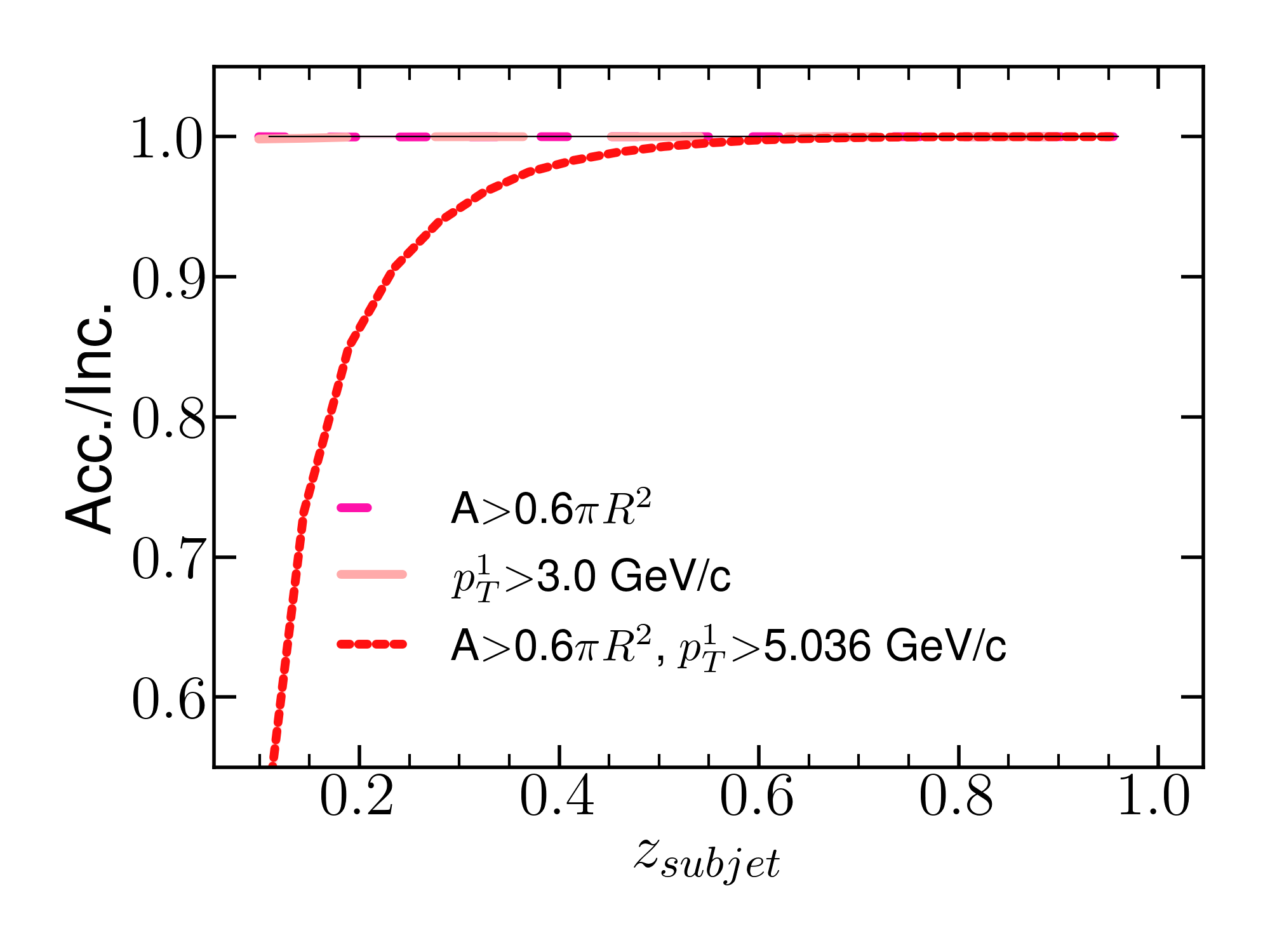}
    \caption{\zSub ratio of accepted signal jets to inclusive signal jets with $R=0.5$ and \ptH$=40$ \GeV after applying each selection. Jets with lower values of \zSub have been suggested to be more gluon-like.}
    \label{fig:z_subjet}
\end{figure}

The selection chosen by the machine learning algorithm suppresses combinatorial jets much more effectively, removing 99.9\% of combinatorial jets. However, signal jets may be more susceptible to survivor bias, as this selection removes 1.6\% of signal jets.  \Fref{fig:z_subjet} shows the distribution of \zSub for each kinematic selection compared to the inclusive distribution for signal jets.  The selection on $A$ and \pTone $> 3.0$ \GeV are consistent with one for all \zSub.  For the kinematic selection chosen by the machine learning algorithm, low \zSub signal jets are significantly suppressed. This suppression is more pronounced for jets with larger resolution parameters and lower \ptH, shown in Supplemental Material~\cite{[{See Supplemental Material at }][{ for kinematic selections of other values of R, and \ptH.}]supp}.

\section{Conclusions}
We used the background generator \BG combined with signal jets from PYTHIA to investigate ways that signal and combinatorial jets can be distinguished in heavy ion collisions. \BG events only contain correlations due to flow, providing an unambiguous definition of the combinatorial background. We use properties which could be reproduced in data, $A$, \pTone, \jetW, and \meanpT, to describe each jet.  We find that signal and combinatorial jets overlap inextricably.  The silhouette values show that they are indistinguishable using properties which could realistically be used in data.  Any kinematic selection to reduce the number of combinatorial jets leaves a population of imposter jets which look like signal jets.  

Most kinematic selections to reduce combinatorial jets, aside from a loose cut on the area, reduce the population of signal jets as well.  While a loose selection on \pTone does not appear to impose as much bias, the tighter selection suggested by the machine learning system significantly biases the surviving jet population towards quark-like jets.
If such a selection were applied in data, this indicates that the surviving signal jets are biased towards quark-like jets.  Measurements where corrections for kinematic selections are made by an unmodified simulation, such as PYTHIA, could be correcting for unmeasured gluon-like jets with measurements of quark-like jets. Complicated methods for distinguishing signal and combinatorial jets, such as black-box machine learning or correcting for imposter jets using unfolding, may have model-dependent assumptions.  The possibility of such issues should be clearly elucidated in studies which use these methods.  We call for a greater focus on any assumptions made when subtracting combinatorial background and biases introduced by methods for suppressing and subtracting this background. 
\section{Acknowledgements}
We are grateful to Mateusz Ploskon, Raghav Elayavalli, and Hannah Bossi for feedback on the manuscript.  This work was supported in part by funding from the Division of Nuclear Physics of the U.S. Department of Energy under Grant No. DE-FG02-96ER40982.
We also acknowledge support from the UTK and ORNL Joint Institute for Computational Sciences Advanced Computing Facility.

\bibliography{apssamp}
\onecolumngrid
\newcommand{\doPlotAppendix}{True}

\ifthenelse{\equal{\isArxiv}{True}}{
\newcommand{\buildAppendix}{
\clearpage 
\appendix
\ifthenelse{\equal{\doPlotAppendix}{True}}{
\newcommand{\makePlotAppendix}{
\begin{center}
Appendix A: Figures for all $p_T^{hard min.}$ and R
\end{center}
\newcommand{\makePlotsForRptm}[2]{
\subsection{R=0.#1, $p_T^{hard min.}$=#2 GeV/c}
\begin{figure}[h]
    \centering
    \includegraphics[width=1.05\linewidth]{Appendix_Plots/blob_plots_fixed_0#1_#2.png}
    \caption{Jet area vs. jet width for R=0.#1 \ptH=#2 \GeV.}
    \label{fig:blob_0#1_#2}
\end{figure}
\begin{figure}[h]
    \centering
    \includegraphics[width=1.05\linewidth]{Appendix_Plots/inc_for_area_cut_fixed_0#1_#2.png}
    \caption{Jet a) leading hadron momentum, b) width, and c) mean constituent momentum for R=0.#1 \ptH=#2 \GeV after area selection.}
    \label{fig:area_0#1_#2}
\end{figure}
\begin{figure}[h]
    \centering
    \includegraphics[width=1.05\linewidth]{Appendix_Plots/inc_for_ml_cut_fixed_0#1_#2.png}
    \caption{Jet a) area, b) width, and c) mean constituent momentum for R=0.#1 \ptH=#2 \GeV after area selection and ML suggested stricter leading hadron selection.}
    \label{fig:ml_0#1_#2}
\end{figure}
\begin{figure}[h]
    \centering
    \includegraphics[width=1.05\linewidth]{Appendix_Plots/inc_for_pT1_cut_fixed_0#1_#2.png}
    \caption{Jet a) area, b) width, and c) mean constituent momentum for R=0.#1 \ptH=#2 \GeV after leading hadron selection.}
    \label{fig:pT1_0#1_#2}
\end{figure}
\begin{figure}[h]
    \centering
    \includegraphics[width=1.05\linewidth]{Appendix_Plots/silhouette_values_fixed_0#1_#2.png}
    \caption{Silhouette values for each kinematic selection. a) Inclusive, b) area selection, c) leading hadron momentum selection, d) area+stricter leading hadron momentum selection for R=0.#1 \ptH=#2 \GeV.}
    \label{fig:sil_0#1_#2}
\end{figure}
\begin{figure}[h]
    \centering
    \includegraphics[width=0.5\linewidth]{Appendix_Plots/z_subjet_ratio_0#1_#2.png}
    \caption{$z_{subjet}$ ratio of accepted signal jets for each kinematic selection to the inclusive signal jet population for R=0.#1 \ptH=#2 \GeV.}
    \label{fig:z_sub_0#1_#2}
\end{figure}
\clearpage
}

\makePlotsForRptm{2}{10}
\makePlotsForRptm{2}{20}
\makePlotsForRptm{2}{30}
\makePlotsForRptm{2}{40}
\makePlotsForRptm{2}{60}
\makePlotsForRptm{2}{80}

\makePlotsForRptm{3}{10}
\makePlotsForRptm{3}{20}
\makePlotsForRptm{3}{30}
\makePlotsForRptm{3}{40}
\makePlotsForRptm{3}{60}
\makePlotsForRptm{3}{80}

\makePlotsForRptm{4}{10}
\makePlotsForRptm{4}{20}
\makePlotsForRptm{4}{30}
\makePlotsForRptm{4}{40}
\makePlotsForRptm{4}{60}
\makePlotsForRptm{4}{80}

\makePlotsForRptm{5}{10}
\makePlotsForRptm{5}{20}
\makePlotsForRptm{5}{30}
\makePlotsForRptm{5}{40}
\makePlotsForRptm{5}{60}
\makePlotsForRptm{5}{80}

\makePlotsForRptm{6}{10}
\makePlotsForRptm{6}{20}
\makePlotsForRptm{6}{30}
\makePlotsForRptm{6}{40}
\makePlotsForRptm{6}{60}
\makePlotsForRptm{6}{80}

}
}{
\newcommand{\makePlotAppendix}{}
}
\makePlotAppendix

\begin{center}
Appendix B: Kinematic selections extracted from machine learning system
\end{center}
\renewcommand{\arraystretch}{2}
\begin{table*}[h]
    \centering
    \begin{tabular}{|c|c|c|c|p{.07\textwidth}|p{.07\textwidth}|p{.07\textwidth}||p{.07\textwidth}|p{.07\textwidth}|p{.07\textwidth}||p{.07\textwidth}|p{.07\textwidth}|p{.07\textwidth}|}
    \hline
R & \ptH & Feature & Selection Threshold & \% Comb. Jets Rejected for ML & \% Signal Jets Rejected for ML & S/B Increase & \% Comb. Jets Rejected for ML + Area Selection  & \% Signal Jets Rejected for ML + Area Selection & S/B Increase & \% Comb. Jets Rejected for \pTone selection  & \% Signal Jets Rejected for \pTone selection & S/B Increase \\
\hline
0.2 & 10 \GeV & \pTone & $>$3.349 \GeV & 98.01\% & 5.89\% & 47.3 & 98.11\% & 6.03\% & 49.7 & 95.73\% & 2.64\% & 22.8\\
 & 20 \GeV & \pTone & $>$4.021 \GeV & 99.50\% & 1.03\% & 197.9 & 99.52\% & 1.04\% & 206.2 & 95.66\% & 0.04\% & 23.0\\
 & 40 \GeV & \meanpT & $>$0.862 \GeV & 98.24\% & 0.24\% & 56.7 & 99.95\% & 0.24\% & 1995.2 & 95.69\% & 0.00\% & 23.2\\
 & 80 \GeV & \meanpT & $>$1.122 \GeV & 99.49\% & 0.03\% & 196.0 & 100\% & 0.03\% & $\infty$ & 95.75\% & 0.00\% & 23.5\\\hline
0.3 & 10 \GeV & \pTone & $>$3.394 \GeV & 96.21\% & 15.05\% & 22.4 & 96.59\% & 15.54\% & 24.8 & 91.38\% & 7.24\% & 10.8\\
 & 20 \GeV & \pTone & $>$3.952 \GeV & 98.92\% & 3.20\% & 82.0 & 99.00\% & 3.21\% & 96.8 & 91.61\% & 0.23\% & 11.9\\
 & 40 \GeV & \pTone & $>$5.257 \GeV & 99.96\% & 0.83\% & 2479.3 & 99.96\% & 0.83\% & 2479.3 & 91.81\% & 0.00\% & 12.0\\
 & 80 \GeV & \meanpT & $>$0.825 \GeV & 97.94\% & 0.08\% & 48.5 & 99.99\% & 0.08\% & 9992.0 & 91.69\% & 0.00\% & 12.0\\\hline
0.4 & 10 \GeV & \pTone & $>$3.486 \GeV & 95.07\% & 25.92\% & 15.0 & 95.74\% & 26.91\% & 17.2 & 86.51\% & 11.85\% & 6.5\\
 & 20 \GeV & \pTone & $>$3.975 \GeV & 98.27\% & 6.59\% & 54.0 & 98.44\% & 6.65\% & 59.8 & 86.67\% & 0.51\% & 7.5\\
 & 40 \GeV & \pTone & $>$5.125 \GeV & 99.86\% & 1.56\% & 703.1 & 99.88\% & 1.56\% & 820.3 & 86.89\% & 0.00\% & 7.6\\
 & 80 \GeV & \meanpT & $>$0.706 \GeV & 96.59\% & 0.25\% & 29.3 & 99.98\% & 0.25\% & 4987.5 & 87.04\% & 0.00\% & 7.7\\\hline
0.5 & 10 \GeV & \pTone & $>$3.497 \GeV & 92.83\% & 33.01\% & 9.3 & 94.09\% & 34.50\% & 11.1 & 81.41\% & 14.77\% & 4.6\\
 & 20 \GeV & \pTone & $>$4.082 \GeV & 97.92\% & 11.07\% & 42.8 & 98.26\% & 11.24\% & 51.0 & 81.21\% & 0.94\% & 5.3\\
 & 40 \GeV & \pTone & $>$5.036 \GeV & 99.79\% & 2.23\% & 465.6 & 99.81\% & 2.23\% & 514.6 & 81.23\% & 0.00\% & 5.3\\
 & 80 \GeV & \meanpT & $>$0.647 \GeV & 95.04\% & 0.79\% & 20.0 & 99.96\% & 0.79\% & 2480.3 & 81.72\% & 0.00\% & 5.5\\\hline
0.6 & 10 \GeV & \pTone & $>$3.424 \GeV & 88.88\% & 33.45\% & 6.0 & 91.17\% & 35.16\% & 7.3 & 75.95\% & 15.51\% & 3.5\\
 & 20 \GeV & \pTone & $>$4.178 \GeV & 97.79\% & 15.54\% & 38.2 & 98.14\% & 15.79\% & 45.3 & 76.41\% & 1.17\% & 4.2\\
 & 40 \GeV & \pTone & $>$5.222 \GeV & 99.77\% & 3.83\% & 418.1 & 99.79\% & 3.83\% & 458.0 & 76.40\% & 0.00\% & 4.2\\
 & 80 \GeV & \meanpT & $>$0.612 \GeV & 93.24\% & 1.40\% & 14.6 & 99.87\% & 1.40\% & 758.5 & 76.62\% & 0.00\% & 4.3\\\hline
\end{tabular}
    \caption{Kinematic selections determined by the machine learning algorithm described in \sref{Sec:MachineLearning}, along with the rejection rates for signal and combinatorial jets.}
    \label{tab:MLcuts}
\end{table*}
\clearpage
\renewcommand{\arraystretch}{1}
}
}{
\newcommand{\buildAppendix}{}
}
\buildAppendix

\end{document}